# Risk Assessment For Spreadsheet Developments: Choosing Which Models to Audit

Raymond J. Butler, CISA
H. M. Customs and Excise, UK
Email ray.butler@hmce.gov.uk

**ABSTRACT**

*Errors in spreadsheet applications and models are alarmingly common (some authorities, with justification cite spreadsheets containing errors as the norm rather than the exception). Faced with this body of evidence, the auditor can be faced with a huge task - the temptation may be to launch code inspections for every spreadsheet in an organisation. This can be very expensive and time-consuming. This paper describes risk assessment based on the "SpACE" audit methodology used by H M Customs & Excise's tax inspectors. This allows the auditor to target resources on the spreadsheets posing the highest risk of error, and justify the deployment of those resources to managers and clients. Since the opposite of audit risk is audit assurance the paper also offers an overview of some elements of good practice in the use of spreadsheets in business.*

## 1. INTRODUCTION

There is a huge body of evidence (summarised in Panko, 2000[1] and Creely, 2000[2]) that errors in spreadsheet applications and models are alarmingly common (some authorities, with justification cite spreadsheets containing errors as the norm rather than the exception)

Spreadsheet users, developers, and auditors need to define controls in order to manage and reduce this risk.

**Preventive Controls**

Sound development methods, standards and user education are the obvious preventive controls, These have been described for almost as long as electronic spreadsheets have been available (e.g. Nevison, 1987[3] and Batson & Eyles, 1995[4])

The many examples of errors found in both field audits and experiments (Panko, 2000[1], Butler, 2000[5]) and studies such as that of Galletta & Hufflagel, 1992[6] show that:

- good development practice is rarely codified into business procedures, and

- Even when it is, the rules and restrictions it requires are not followed to any significant degree.

**Detective and Corrective Control**

Detective and corrective controls over spreadsheet errors are principally provided by detailed code inspection to check the formulas and where necessary the input data. This
can be

- an entirely manual process, or

- One of the several computer-assisted audit tools for spreadsheets may be used.

Code inspection may be performed by

- one individual (not particularly effective - Galletta, Abraham, Louadi, Lekse, Pollailis and Sampler, 1993 [7] found it to be less than satisfactory, detecting only around 50% of errors)

- by teams of 2 or more (More effective, finding around 80% of errors - Panko, 1996[8]).

**The Auditor's Problem**

Faced with this body of evidence, the auditor can be faced with a huge task – the temptation may be to perform a full code inspection on every spreadsheet encountered. However, effective code inspection, even when aided by Computer-assisted audit tools, can be extremely resource intensive. Given limited resources, even with the anecdotal evidence of the inevitability of errors in spreadsheet model development, auditors must prioritise their work and justify the expense of code inspection to management and clients have to determine

- the potential impact of errors in a model on the organisation *and*

- The likely incidence of errors in the model.

This paper describes a risk assessment methodology used (with some success) by officers of H M Customs & Excise to determine which spreadsheet applications they need to test in depth in order to gain assurance that they calculate taxes and duties correctly.

**A note on Terminology**

Theterminology used in this paper reflects the fact that Microsoft Excel is the spreadsheet programme used by H M Customs & Excise. However, the concepts and methods set out below apply to any electronic spreadsheet programme and (with minor modifications) to other areas of end-user computing. In this paper

- *spreadsheet is* used to mean the electronic spreadsheet program (e.g. Microsoft *Excel,* Lotus 1-2-3) used to develop a

- *model* - a generally complex single use development for financial or other planning purposes or an

- *Application,* which may be simple or complex but is generally used as a regular part of a business' operation.

**2. THE RISK ASSESSMENT**

Themethodology described is a multi-stage process, which allows the auditor to make a "stop / go" decision at each stage Figure 1 illustrates this. The first two stages of Risk Assessment considers the environment within which the model or application is developed. The subsequent steps move on to consider the development itself.

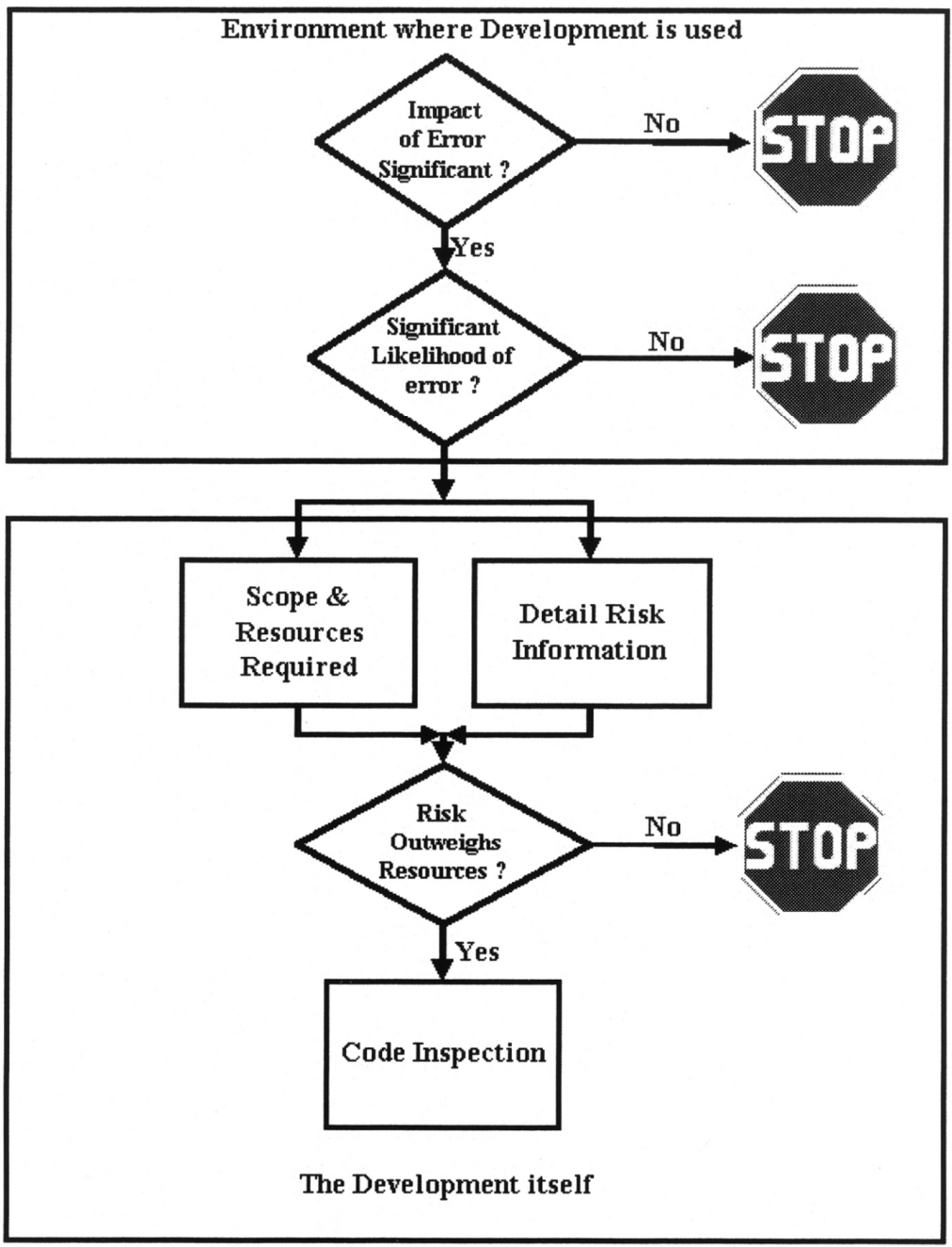

**Figure 1 - Overview of Risk Assessment Methodology**

## 2.1 OVERALL RISK ASSESSMENT

All audit planning is governed by the need to deploy resources to address quantified risks. The auditor must therefore determine both the impact and possible incidence of risks before setting out on a testing programme.

The first two stages of risk assessment - the overall assessment - are performed without looking at the detail of the model or application at all. All the steps, and the evaluation, may be taken relatively quickly, and the time saved by avoiding unnecessary code inspections amply justifies the effort of risk assessment.

**Impact of Errors**

The auditor must determine the:

- amount of money or other resources handled by or at risk from the application or model
- regulatory consequences of any errors
- impact of any errors on the organisation's public image, shareholder etc. confidence

The span of this will cover either the current instance or (if the current instance is a template for future re-use) a year or some other convenient period. The re - use of an application as a template for business operations will multiply both the resources handled and the impact of any errors. At this point the auditor can decide whether the amount at risk from the application justifies the work required to ascertain the likely incidence of errors.

**Likely Incidence of Errors**

When the Auditor has that the impact of an error is likely to be significant, the auditor must make a judgement as to the likelihood of that exposure occurring. To inform this judgement, the auditor must consider the answers to the following questions:

**Organisation Questions**

Does the organisation for whom the development is being made have an adequate policy governing development, testing and use of spreadsheet models and applications?

What evidence is there that this policy is observed and enforced?

**Domain Questions**

How complex are the business or revenue issues that the model or application purports to address?

Is there evidence that the developer of the model or application has

- An adequate understanding of those issues?
- Access to a clear, accurate, written explanation of the business issues?

If domain knowledge is partial or absent, errors in the base calculations or of omission are much more likely. Further, it follows that the developer will be less able to detect errors through a "reasonableness test" of the output from a model or application.

**Specification Questions**

- Is there a clear statement of the inputs, processes, outputs and results required for the
- model or application?
- Is it complete and accurate?
- Has the user agreed it?

- Does it include agreed measures of the success of the development?

- Does it include a testing plan?

If no specification exists, then domain and arithmetical errors are much more likely, and there will be no control against which the completeness and accuracy of the results can be judged. Developing a complex spreadsheet application or model without a specification is akin to walking across a swamp without a map - the only measure of success is surviving the experience.

**Testing Questions**

What evidence is there that the application was thoroughly tested before being brought into use? And thoroughly tested again each time a material change was made? If a model or application can be shown to have given

- sensible answers when tested with simple numbers (for example, if the answer to =+5+5 is 7 then there's a problem somewhere) and

- the results predicted when realistic test data are processed and

- the results predicted when running the model in parallel with previous systems

Then there is a good level of assurance that the incidence of errors will be low.

**Documentation Questions**

Has the developer documented the application adequately? Good documentation should make clear statements of:

- the application's purpose, what it does and how it does it

- any assumptions made in its design

- what standing data constants (e.g. tax, duty, interest and exchange rates) are used and where they are held

- who developed it and when, and

- When and how it has been changed since being brought into use.

- How the application or model should be used.

The absence of documentation has been a factor in a number of well-documented spreadsheet errors. Where a developer is not the end-user of an application, where any application is more than utterly simple, and wherever the developer will not be maintaining and updating the application documentation is absolutely essential. Failure to document the inner workings of a model simply stores up trouble for whoever has to amend or maintain it. Failure to document its correct operation by the end-users can lead to serious errors, especially in an environment where a model or application is passed from user to user.

**Questions about the complexity of the application**

How complicated is the application? Is it laid out logically, with data and calculations in separate areas, and with complex calculations broken down into stages? Is the arithmetic of the calculations clear from the visible information, allowing it to be checked manually for accuracy and completeness?

**Data Control Questions**

In common with all computer applications, the accuracy of the results of processing depend on the completeness, accuracy, timeliness and authorisation/appropriateness of the data. Even when a properly validated specification has been verified as implemented correctly, with all domain and arithmetic issues correctly handled, the GIGO (for younger readers, Garbage in, Garbage Out) principle still applies.

The auditor must therefore ask what controls are built into the application to ensure that:

- all relevant data are input,
- no irrelevant or inappropriate data are input
- data are input accurately, and
- data are input for process at the correct time

**Evaluation**

The auditor must consider the answers to these questions to inform a decision whether or not it is worth proceeding to the next stage. It is important that the *quality* of documentation, test plans, etc is taken into account in this decision - good practice in these areas is very rare, and auditors must beware of being led to a false sense of security by the very *existence* of documentation, user instructions, etc.

Given that

- the amount of resource potentially at risk, and
- the evaluation of the above factors

Shows that the impact and likelihood of errors justifies further work, the auditor can progress to the next stage of risk assessment. This stage determines whether a full code inspection may be required, how much effort may be involved in it, and whether the effort will be justified by the risk.

**2.2 RISK IDENTIFICATION & SCOPING**

Given that the impact and likelihood of error justify further work, the auditor now needs to establish

- the size and complexity of the application (to help plan the time needed to test it),
- Which parts of it pose the highest risk (to help direct the tests to those risk areas).

This step requires access to the model or application, since its composition and set-up have to be assessed in order to inform the stop - or- go decision for code inspection. It is greatly eased by the use of spreadsheet audit software, which can quickly reveal the inner workings of an application.

**Size of the task**

To determine the complexity of the checking task, and the amount of time that may be needed to perform a code inspection, the auditor needs to know:

- How many physical files are involved in the application or model?

- for each file, how many worksheets are present

- if data is passed from one file or worksheet to another, are adequate controls in place to ensure the completeness and accuracy of the transfer?

Transposition of digits and other keying errors can easily corrupt the receiving file or worksheet. Similarly, automatic links between files must be subjected to controlled for completeness, accuracy and appropriateness of data transfer. This allows the auditor to determine the boundaries of the audit. Within these boundaries, the auditor then needs to establish for each worksheet within each file:

- how many formulas are present,

- how many numbers are manipulated,

- how many labels are present and

- How many links to other worksheets exists.

Many of the audit support software products allow this information to be produced automatically. This information is used to inform decisions on

- time management (how much time is code inspection likely to take?) and

- Resource to risk (does the money at risk justify spending that time?).

**How complex is the task?**

Assuming the auditor has justified expending further resources, a better indication of the time that has to be expended on testing is needed. The auditor now has to establish how many

- external references,

- unique formulas (i.e. those which are not replicated in a worksheet), and

- original formulas (i.e. those which are copied within a worksheet)

Are present in each file. The degree to which similar worksheets are used within a file or across a series of (ostensibly) identical files is also a factor in determining the resources needed for a code inspection. Once a master original worksheet has been tested and if necessary corrected and documented, to establish a norm for the audit, automated comparison of worksheets can quickly identify any divergences from that norm in copies. This can significantly reduce theamount of actual code inspection testing that has to be performed.

Theauditor has to consider how complex the business issues the application addresses are,and how complex its structure and logic are. Drawing a map or flowchart of the application at this point can help comprehension of the structure and interaction of its components

**Identification of Set-up Risks**

Identification of the use of high-risk functions or features, and an assessment of the use of security features and the way the application is set up helps the auditor to judge the amount of risk and the amount of work needed to test the application. The auditor needs to establish

- The recalculation settings of each file

  - Manual or automatic?

  - If manual, is it set to recalculation before save set?

  - How are iteration & calculation rules set?

- Whether macros and user-defined functions are present, or if indications are present of any traces of their use - with modem spreadsheets, macros are often an attribute of the user's individual set-up rather than the spreadsheet file itself,

  - Whether Hidden Rows, Columns or sheets are present in the file,

  - Whether protection against unauthorised changes is present

  - Whether advanced features such as consolidation, scenarios, goal seeking, solver, pivot tables, report or view manager and equivalent features are used.

If the results of a model or application depend on the use of these features, the auditor will have to consider whether the techniques are appropriate to resolution of the issues being dealt with and whether they are being used correctly.

  - Whether range and variable names are being used in formulas - this can indicate a developer's use of good practice. However, the auditor will have to consider whether names are being used correctly.

**2.3 THE TESTING DECISION**

At this stage, the auditor will know:

- the amount at risk and the likely incidence of risk from the model or application

- the amount of effort that is likely to be required in order to manage that risk by substantive testing

- whether the balance of the amount and incidence of risk justifies that further work, and (as a by-product of the risk assessment/compliance testing); and

- Which areas of the application may require detailed scrutiny

## 2.4 RISK IN THE CODE INSPECTION PHASE

**Code Inspection**

Code Inspection will, if supported by adequate software, be targeted on risks that –

- Original formulas copied around the worksheet or workbook are arithmetically and or logically incorrect.

- Copies of those formulas are used inappropriately.

- Unique formulas are arithmetically and logically incorrect.

- Formulas have been over-written by numbers or other data.

**Additional Risks in the Code Inspection**

Even if they appear arithmetically and logically correct, further checks will need to be made on formulas that present a high risk of error, i.e. those which

- look up named ranges, e.g. standing data;

- contain constants (e.g. net * 17.5% instead of net * a named variable or range "VAT rate");

- contain absolute references (which will not automatically respond to changes to the sheet);

- reference a block of cells (e.g. SUM Al:B7 may indicate errors in input of the formula);

- Have no precedents (e.g. additions of numbers within a cell, which invariably gives rise to audit trail problems).

Depend on

- numbers formatted as text (which may cause errors or unpredictable results), or

- blank cells (which may reveal errors in construction or in data input),

- 
- have no dependant cells (if not the end result, may be an error),

- address hidden cells, rows or columns,

- address cells which fail or return an error message,

- address linked sheets and workbooks,

- have an inherently high risk of user error (e.g. =NPV) or

- show as an apparent break in the pattern of formulas copied from a single source -

Most test support software will help the auditor to identify these.

**Data Checking**

Given the high risk of errors in formulas, it can be easy to overlook the issues of errors in the data - as stated above, the GIGO principle applies to spreadsheets as much as to any computer application. The auditor will already have risk analysed the procedures and controls over data, and depending on the presence (or absence) and quality of systems in place to assure the completeness and accuracy of data may need to substantively test the base numbers and standing data used. In particular, the auditor must consider the risks that:

- incorrect or inappropriate raw numbers could be introduced into the model;

- numbers may be incorrectly input in place of formulas;

- numbers may have been introduced and not used by any formula; and

- numbers may have been incorrectly formatted as text, leading to their omission from totals, etc.

All these circumstances are known to pose a high risk of error.

## 3. CONCLUSIONS

Risk assessment is at the heart of all auditing, whether of manual accounts, complex enterprise resource planning systems, or of spreadsheets.

The risks of error arising from poor practice in the use of spreadsheets are known to be high, and the incidence of good practice in developments using spreadsheets are known to be low. Despite this, users are blissfully unaware of these risks and are using potentially faulty decision support machinery every day to take vital business decisions.

Given this, and that an error in a spreadsheet application can subvert all the controls in all of the systems which feed data into it, risk assessment in spreadsheets is vital if good practice is to be encouraged, bad practice detected and corrected, and appropriate resources put into auditing by data and code inspection.

This paper is offered as a starter for further research into the effectiveness of error prevention and detection methods and to inform future audits by IS and other auditors.

# REFERENCES

1. Panko, R. (1996) Hitting the Wall: Errors in developing & debugging a "simple" Spreadsheet model" Proceedings of the 29th Hawaii International Conference on System Sciences

2. Creely, Paul "Review of Errors in Spreadsheets", *Final Year BSc Paper, University of Salford* 2000

3. Nevinson, J. M. (1987)"The Elements of Spreadsheet Style" Brady / Prentice Hall,

4. Batson, J. Eyles, J (1995) "Spreadsheet Modelling Best Practice" Accountants Digest, April 1995

5. Butler, R J 'Is This Spreadsheet a Tax Evader? How H M Customs & Excise test Spreadsheet Applications" *Proceedings of the 33rd Hawaii International Conference on System Sciences, Maui, Hawaii,* 2000

6. Galletta, D. F., & Hufnagel, E. M. (1992). A Model of End-User Computing Policy: Context, Process, Content and Compliance. Information and Management, 22(1), 1-28.(Cited in Creely 2000)

7. Galletta, D.F.; Abraham, D.; EI Louadi, M.; Lekse, W.; Pollailis, Y.A.; & Sampler, LL. "An Empirical Study of Spreadsheet Error-Finding Performance," Journal of Accounting, Management, and Information Technology (12) April-June 1993, pp. 79-95. (Cited in Creely, 2000)

8. Panko, R (ed.) Spreadsheet Research (SSR) Web Site http://www.cba.hawaii.edu/panko/ssr,2000